\documentclass{emulateapj}
\usepackage{apjfonts}
\slugcomment{Accepted by  \textsc{the Astrophysical Journal}}

\newcommand{\dif}{\ensuremath{\mathrm{d}}}

\newcommand{\ee}[1]{\ensuremath{\times 10^{#1}}}

\newcommand{\kB}{\ensuremath{k_\mathrm{B}}} 
\newcommand{\mb}{\ensuremath{m_\mathrm{u}}} 
\newcommand{\me}{\ensuremath{m_\mathrm{e}}} 

\newcommand{\unitspace}{\ensuremath{\,}}
\newcommand{\usp}{\unitspace}
\newcommand{\numberspace}{\ensuremath{\;}}
\newcommand{\nsp}{\numberspace}
\newcommand{\unitstyle}[1]{\ensuremath{\mathrm{#1}}}
\newcommand{\power}[2]{\ensuremath{{#1}^{#2}}}

\newcommand{\centi}{\unitstyle{c}}
\newcommand{\kilo}{\unitstyle{k}}
\newcommand{\Mega}{\unitstyle{M}}

\newcommand{\meter}{\unitstyle{m}}

\newcommand{\second}{\unitstyle{s}}

\newcommand{\Kelvin}{\unitstyle{K}}
\newcommand{\K}{\Kelvin}  

\newcommand{\cm}{\centi\meter}
\newcommand{\gram}{\unitstyle{g}}

\newcommand{\grampercc}{\gram\usp\power{\cm}{-3}} 
\newcommand{\grampersquarecm}{\gram\usp\power{\cm}{-2}} 

\newcommand{\columnunit}{\grampersquarecm}

\newcommand{\eV}{\unitstyle{eV}}        
\newcommand{\keV}{\kilo\eV} 
\newcommand{\MeV}{\Mega\eV} 

\newcommand{\Msun}{\ensuremath{M_\odot}}

\newcommand{\nuclei}[2]{\ensuremath{\mathrm{^{#1}#2}}}
\newcommand{\carbon}[1][12]{\nuclei{#1}{C}}
\newcommand{\oxygen}[1][16]{\nuclei{#1}{O}}
\newcommand{\magnesium}[1][24]{\nuclei{#1}{Mg}}
\newcommand{\aluminum}[1][27]{\nuclei{#1}{Al}}
\newcommand{\iron}[1][56]{\nuclei{#1}{Fe}}
\newcommand{\ironfiftyfour}[1][54]{\nuclei{#1}{Fe}}
\newcommand{\nickelfiftyeight}[1][58]{\nuclei{#1}{Ni}}

\newcommand{\igncol}{\ensuremath{\Sigma_{\mathrm{ign}}}}
\newcommand{\source}[3]{#1~#2$#3$}
\newcommand{\ks}{\source{KS}{1731}{-260}}
\newcommand{\mxb}{\source{MXB}{1659}{-29}}
\newcommand{\lmxbtrans}{\source{4U}{1608}{-522}}
\newcommand{\sdot}{\dot{\Sigma}}
\newcommand{\sdotEdd}{\sdot_{\mathrm{Edd}}}

\newcommand{\pd}[2]{\frac{\partial #1}{\partial #2}}
\newcommand{\estar}{\mathcal{E}}
\newcommand{\Qstar}{\mathcal{Q}}
\newcommand{\chempot}{\mu_{\mathrm{e}}}
\newcommand{\ecrate}{r_{\mathrm{ec}}}
\newcommand{\nucrate}{r_{\mathrm{nuc}}}
\newcommand{\ft}{{\langle f \! t \rangle}}
\newcommand{\epsnuc}{\epsilon_{\mathrm{nuc}}}

\newcommand{\epsnu}{\epsilon_{\nu}}
\newcommand{\tacc}{t_{\mathrm{acc}}}
\newcommand{\Sigmaec}{\Sigma_{\mathrm{ec}}}
\newcommand{\Tec}{T_{\mathrm{ec}}}
\newcommand{\rtp}{r_{\mathrm{TP}}}
\newcommand{\Epk}{E^{\mathrm{pk}}}
\newcommand{\Fex}{F^{\mathrm{ex}}}
\newcommand{\fex}{f^{\mathrm{ex}}}
\newcommand{\delfex}{\Delta \fex}
\newcommand{\sigmav}{\langle \sigma v \rangle}
\newcommand{\ER}{E_{\mathrm{R}}}
\newcommand{\og}{(\omega \gamma)_{\mathrm{R}}}
\newcommand{\GammaC}{\Gamma_{\mathrm{C}}}
\newcommand{\GammaR}{\Gamma_{\mathrm{R}}}
\newcommand{\SR}{S_{\mathrm{R}} (E)}
\newcommand{\Sexp}{S_{\mathrm{exp}}(E)}
\newcommand{\Teight}{T_{8}}
\newcommand{\Stwelve}{\Sigma_{12}}
\newcommand{\Qimp}{Q_{\mathrm{imp}}}
\newcommand{\pico}{\unitstyle{p}}
\newcommand{\barn}{\unitstyle{b}}
\newcommand{\avgz}{\langle Z \rangle}
\newcommand{\avgzsq}{\langle Z^{2} \rangle}
\newcommand{\avgzfh}{\langle Z^{5/2} \rangle}

\newcommand{\avgzk}{\langle Z^{k} \rangle}
\newcommand{\dd}{\mathcal{D}}
\newcommand{\zetadh}{\zeta^{\mathrm{DH}}}
\newcommand{\zetalm}{\zeta^{\mathrm{LM}}}

\begin{document}
\title{Possible Resonances in the \carbon\ + \carbon\ Fusion Rate and Superburst Ignition}
\author{Randall L. Cooper}
\affil{Kavli Institute for Theoretical Physics, University of California, Santa Barbara, CA 93106}
\author{Andrew W. Steiner and Edward F. Brown}
\affil{Department of Physics \& Astronomy, National
Superconducting Cyclotron Laboratory, and the Joint Institute for
Nuclear Astrophysics, Michigan State University, East  Lansing, MI}
\email{rcooper@kitp.ucsb.edu; steinera@pa.msu.edu; ebrown@pa.msu.edu}

\begin{abstract}
Observationally inferred superburst ignition depths are shallower than models predict.  We address this discrepancy by reexamining the superburst trigger mechanism.  We first explore the hypothesis of Kuulkers et al.\ that exothermic electron captures trigger superbursts.  We find that all electron capture reactions are thermally stable in accreting neutron star oceans and thus are not a viable trigger mechanism.  Fusion reactions other than \carbon\ + \carbon\ are infeasible as well since the possible reactants either deplete at much shallower depths or have prohibitively large Coulomb barriers.  Thus we confirm the proposal of Cumming \& Bildsten and Strohmayer \& Brown that \carbon\ + \carbon\ triggers superbursts.  We then examine the \carbon\ + \carbon\ fusion rate.  The reaction cross-section is experimentally unknown at astrophysically relevant energies, but resonances exist in the \carbon\ + \carbon\ system throughout the entire measured energy range.  Thus it is likely, and in fact has been predicted, that a resonance exists near the Gamow peak energy $\Epk \approx 1.5 \nsp \MeV$.  For such a hypothetical $1.5 \nsp \MeV$ resonance, we derive both a fiducial value and upper limit to the resonance strength $\og$ and find that such a resonance could decrease the theoretically predicted superburst ignition depth by up to a factor of $4$; in this case, observationally inferred superburst ignition depths would accord with model predictions for a range of plausible neutron star parameters.  Said differently, such a resonance would decrease the temperature required for unstable \carbon\ ignition at a column depth $10^{12}\nsp\columnunit$ from $6\ee{8} \nsp \K$ to $5\ee{8} \nsp \K$.  A resonance at $1.5 \nsp \MeV$ would not strongly affect the ignition density of Type Ia supernovae, but it would lower the temperature at which \carbon\ ignites in massive post--main-sequence stars.
Determining the existence of a strong resonance in the Gamow window requires measurements of the \carbon\ + \carbon\ cross-section down to a center-of-mass energy near $1.5 \nsp \MeV$, which is within reach of the proposed DUSEL facility.
\end{abstract}


\section{Introduction}\label{s.introduction}

Superbursts are long, energetic, and rare thermonuclear flashes on accreting neutron stars in low-mass X-ray binaries.  Their durations ($\sim$ hours), fluences ($\sim 10^{42}$ ergs), and 
recurrence times ($\sim$ years) distinguish superbursts from their typical hydrogen- and 
helium-triggered counterparts \citep[for reviews, see][]{K04,Cumming2005Superbursts:-A-,SB06}.  As of this writing, astronomers have detected 15 superbursts from 10 sources \citep[][and references therein]{K04,intZCC04,RM05,K05,Ketal08}.  

The proposal \citep{cumming.bildsten:carbon,strohmayer.brown:remarkable} that thermally unstable \carbon\ fusion \citep{WT76,taam78:_nuclear,brown98} triggers superbursts offers a reasonable explanation of their origin.  Cooling model fits to superburst light curves \citep{cumming.macbeth:thermal,Cumming2005Long-Type-I-X-r} as well as observed fluences and recurrence times \citep[e.g.][]{KintZC06} suggest ignition column depths $\igncol \approx 10^{12}\nsp\columnunit$, where $\Sigma \equiv \int \rho\,\dif z$ is the radially integrated density.  Previous superburst ignition models \citep{cumming.bildsten:carbon,strohmayer.brown:remarkable,C03,brown:superburst,cooper.narayan:theoretical,Cumming2005Long-Type-I-X-r,Gupta2006Heating-in-the-} demonstrated that \carbon\ ignites at 
$\Sigma \approx 10^{12}\nsp \columnunit$ only if \carbon\ is abundant and the ocean temperature $T \approx 6\ee{8}\nsp\K$ at that column depth; within existing models of nuclear heating in the neutron star crust, such a large temperature requires an inefficient neutrino emission mechanism in the neutron star core and a low thermal conductivity in the neutron star crust, so that the crust is much hotter than the core.  

Recent observations, simulations, and experiments have exposed three fundamental problems with this scenario. 
First and foremost is the inference that the ocean is in fact too cold for  \carbon\ ignition at the inferred column depth $\igncol \approx 10^{12}\nsp\columnunit$.  This comes from fits \citep{Shternin2007Neutron-star-co,BC09} to the quiescent cooling of the quasi-persistent transient \ks\ \citep{wijnands.ea:xmm_1731,rutledge.ea.01:ks1731,Cackett2006Cooling-of-the-}, a system that also exhibited a superburst \citep{kuulkers.ea:ks1731-superburst}.  The timescale for the quiescent luminosity to decrease suggests that the crust's thermal conductivity is high; as a result, the inner crust temperature remains close to that of the core even during the accretion outburst.  \citet{Cackett2008Cooling-of-the-} reach the 
same conclusion for \mxb\ \citep[see also][]{BC09}.  In fact, molecular dynamics simulation results \citep{Horowitz2007Phase-Separatio,Horowitz2008Thermal-conduct,HB09} suggest that the neutron star crust is arranged in a regular lattice and therefore has a high thermal conductivity. Neither shear-induced viscous heating \citep{Piro2007Turbulent-Mixin,KLi09} nor deep crustal heating due to electron captures, neutron emissions, and pycnonuclear reactions  \citep[e.g.,][]{Haensel2008Models-of-crust,Horowitz2007Fusion-of-neutr,GKM08}  can account for the heat necessary to raise the ocean temperature to the required level \citep[although see][who consider heating in strange stars and hybrid stars, respectively]{Page2005Superbursts-fro,BSKB08}. 

Second, evidence of heavy-ion fusion hindrance at extreme sub-Coulomb-barrier energies \citep{Jetal02,JRBJ07,Jetal08} implies that the cross-section and thereby the $\carbon+\carbon$ reaction rate may be orders of magnitude smaller than that assumed in the aforementioned superburst ignition models.  When included in superburst ignition models, heavy-ion fusion hindrance increases $\igncol$ by at least a factor of $2$ \citep{Gasques2007Implications-of}.

Third, the means by which nuclear burning on the stellar surface produces sufficient quantities of \carbon\ to trigger superbursts is poorly understood.  Superburst models require large \carbon\ mass fractions for ignition \citep{cumming.bildsten:carbon,C03,cooper.narayan:theoretical,Cooper2005On-the-Producti,Cumming2005Long-Type-I-X-r}.  All systems that exhibit superbursts show helium-triggered type I X-ray bursts as well \citep[e.g.][]{Galloway2008Thermonuclear-t}, but theoretical models of such bursts yield \carbon\ mass fractions far smaller than those required for ignition \citep{J78,schatz.aprahamian.ea:endpoint,Schatz2003Nuclear-physics,KHKF04,woosley.heger.ea:models,Fisker2005The-reactions-a,FST08,Peng2006Sedimentation-a,PJMI08}.  
Most systems that exhibit superbursts apparently undergo long periods of stable nuclear burning between successive helium-triggered bursts \citep{KHvdKLM02,.kuulkers.ea:superburst,Ketal08}; stable burning generates much more \carbon\ than unstable burning, but the calculated yield is insufficient to trigger superbursts in all systems, particularly those accreting at a high rate \citep{taam78:_nuclear,schatz99,Schatz2003Nuclear-physics,Cooper2005On-the-Producti,FGWD06}.

Detection of a superburst from the classical transient \lmxbtrans\ \citep{RM05,K05} with $\igncol \approx 10^{12}\nsp\columnunit$ \citep{Ketal08} exacerbates all three problems: (1) the transient's inferred ocean temperature is lower than those of other systems exhibiting superbursts, (2) heavy ion fusion hindrance is greater at lower temperatures, and (3) most of the matter accreted onto the neutron star prior to the observed superburst likely burned during helium-triggered type I X-ray bursts, which current theoretical calculations suggest would generate far less \carbon\ than that required for ignition.  

Reconciling superburst observations with the current theoretical model is impossible.  This motivates both a critical assessment of the current ignition model and a search for alternative ignition mechanisms. 

\citet{kuulkers.ea:ks1731-superburst} proposed such an alternative mechanism.  Electron captures onto protons and the subsequent captures of the resulting neutrons onto heavy nuclei liberate $\approx 7\nsp\MeV/\mb$ \citep{bildsten98}. Prethreshold captures of super-Fermi electrons are very temperature sensitive and therefore could trigger an energetic thermonuclear flash.  An attractive feature of this mechanism is that ignition occurs always at the same electron chemical potential; thus $\igncol$ would be similar for all superbursts, in accord with observations.  Unfortunately, the calculated $\igncol \approx 2\ee{10}\nsp\columnunit$ is much smaller than the inferred superburst $\igncol$, making it an unlikely trigger mechanism. This motivates our investigation of exothermic electron captures onto heavy nuclei, which occur throughout the ocean and crust of an accreting neutron star, including the superburst ignition region \citep{sato79,blaes90:_slowl,haensel90a,haensel.zdunik:nuclear,Haensel2008Models-of-crust,Gupta2006Heating-in-the-}.  In \S \ref{s.ecZ} we determine the thermal stability of electron captures onto heavy nuclei in accreting neutron stars.  We show that instability requires unrealistically large reaction $\Qstar$-values, where $\Qstar$ is the energy released per capture; thus we conclude that electron captures in accreting neutron star oceans are thermally stable.  We then consider the relevance of $\alpha$ captures onto light elements such as \carbon\ in \S \ref{s.light-fusion}.  We find that none of these reactions is a feasible mechanism and thereby confirm the proposal that \carbon\ + \carbon\ triggers superbursts.  

In \S \ref{s.carbon-sigma} we assess whether the \carbon+\carbon\ reaction rate could be much larger than the fiducial rate.  We investigate (\S~\ref{s.screening}) the screening enhancement factor, including a careful evaluation of corrections to the liner mixing rule, and show that uncertainties in the plasma screening enhancement are unlikely to change the $\carbon+\carbon$ reaction rate enough. We then consider the nuclear cross-section. We find that a strong resonance at an energy near $1.5 \nsp \MeV$ in the \carbon\ + \carbon\ system, which theoretical nuclear physics models predict, could increase the reaction rate in the astrophysically relevant temperature range by over two orders of magnitude.  In \S~\ref{s.ignition-column} we show that the existence of such a resonance could decrease the predicted $\igncol$ by a factor $\approx 2$--$4$ and thereby alleviate the discrepancy between superburst models and observations.  We conclude in \S \ref{s.discussion} by discussing the implications of our findings.

\section{Thermal Stability of Electron Captures}\label{s.ecZ}

Consider the accretion-driven compression of a matter element containing a nucleus of mass $M(A,Z)$, where $A$ is the mass number and $Z$ is the proton number.  The degenerate electrons' chemical potential $\chempot$ rises as the nucleus advects to higher pressures.  Eventually $M(A,Z) + \chempot/c^{2}$ exceeds $M(A,Z-1)$ and electron capture becomes energetically favorable.  Such captures often occur in equilibrium and release a negligible amount of energy; however, some captures can heat the ocean in two mutually inclusive ways \citep[e.g.][]{Gupta2006Heating-in-the-}: (1) An electron captures into an exited state of the daughter nucleus if, for example, the daughter nucleus's ground state is forbidden.  The daughter nucleus then radiatively deexcites and thereby heats the ocean.  (2) If the parent nucleus is even-even, then $M(A,Z-1) > M(A,Z-2)$ due to the nuclear pairing energy, and a second electron capture immediately ensues.  The latter, post-threshold electron capture occurs out of equilibrium and thus releases heat.  

\subsection{Governing Equations}

We construct a simple model of the accreted layer  to determine the stability of exothermic electron captures to thermal perturbations.  We assume spherical accretion onto a neutron star of mass $M = 1.4 \nsp M_{\odot}$ and radius $R = 10 \nsp \mathrm{km}$ at an accretion rate per unit area $\sdot$.  
The accreted layer's scale height is much less than $R$, so we set the gravitational acceleration $g = GM/R^{2} (1-2GM/Rc^{2})^{-1/2} = 2.43\ee{14}\nsp \cm\usp\power{\unitstyle{s}}{-2}$ throughout the layer.  The layer is always in hydrostatic equilibrium, so the column depth $\Sigma$ is a good Eulerian coordinate.  To facilitate comparisons between microphysical and observationally inferred quantities, we express microphysical quantities in terms of the macroscopic coordinate $\Sigma$ using the following approximate relation between mass density $\rho$ and $\Sigma$ for relativistic, degenerate electrons\footnote{In this and following expressions, we suppress the scaling with $g$ and evaluate the expressions at $g=2.43\ee{14}\nsp\cm\usp\second^{-2}$.}, 
\begin{equation}\label{e.approx-rho}
\rho \approx 5.9\ee{8}\nsp\grampercc \left ( \frac{\langle A/Z \rangle }{2} \right )
	\Stwelve^{3/4},
\end{equation}
where $\langle A/Z \rangle$ is the mean molecular weight per electron and $\Sigma = \Stwelve \times 10^{12}\nsp \columnunit$.  We denote the Eulerian time and spatial derivatives as $\partial/\partial t$ and $\partial/\partial \Sigma$, respectively, and the Lagrangian derivative following a matter element as $D/Dt$, where $D/Dt = \partial/\partial t + \sdot \partial/\partial \Sigma$.  
The governing transport, entropy, and continuity equations are
\begin{eqnarray}
F &=& \rho K \pd{T}{\Sigma}, \label{e.transport}\\
T\frac{Ds}{Dt} &=& \estar X \ecrate + \pd{F}{\Sigma}, \label{e.entropy}\\
\frac{DX}{Dt} &=& -X \ecrate\label{e.continuity},
\end{eqnarray}
where $F$ is the flux, $K$ is the thermal conductivity, $s$ is the entropy, $\estar = \Qstar/(A\mb)$ is the energy per gram released via electron captures, $X$ is the parent nucleus mass fraction, \begin{equation}
\ecrate = \left (\frac{\ln 2}{\ft} \right ) \frac{1}{(\me c^{2})^{5}} \int^{\infty}_{Q} E^{2} (E-Q)^{2} f(E,\chempot,T)dE
\end{equation}
is the electron capture rate \citep{FFN85}, $\ft$ is the effective $f\!t$ value \citep{FFN80,FFN85,langanke.martinez-pinedo:weak}, $\me$ is the electron mass, $Q$ is the threshold energy, and 
\begin{equation}\label{e.FDdistribution}
f(E,\chempot,T) = \frac{1}{1 + \exp[(E-\chempot)/\kB T]}
\end{equation}
is the Fermi-Dirac distribution function.  

Consider prethreshold electron captures, where $\chempot < Q$.  For $T=0$, all electrons have energies $E \leq \chempot$ by equation (\ref{e.FDdistribution}); electron capture is blocked.  For $T>0$ some electrons have $E>Q$ and thus can capture.  The number of electrons with $E>Q$ increases with $T$, which makes prethreshold electron capture temperature-sensitive.  In the prethreshold limit $(\chempot-Q)/\kB T \ll 0$, 
\begin{equation}\label{e.prethresholdrate}
\ecrate(\chempot,T) = \left (\frac{\ln 2}{\ft} \right ) \frac{2Q^{2}(\kB T)^{3}}{(\me c^{2})^{5}} \exp \left [ - \left (\frac{Q-\chempot}{\kB T} \right ) \right ]
\end{equation}
\citep{FFN85,bildsten98}.  Conversely, for $\chempot > Q$ a majority of electrons has $E > Q $ and hence can capture for any $T$, making $\ecrate$ relatively temperature-insensitive and thus thermally stable.  We therefore consider prethreshold electron captures exclusively hereafter.  

\subsection{Prethreshold Electron Capture}

Prethreshold electron captures occur within a thin layer in the deep ocean.  To illustrate this, consider the height-integrated capture rate.  Relativistic, degenerate electrons supply the pressure $P=g\Sigma$, so
\begin{equation}\label{e.Sigmaofchempot}
\Sigma \approx \frac{\chempot^{4}}{12\pi^{2}g(\hbar c)^{3}} = 10^{12}\nsp\columnunit \left(\frac{\chempot}{3.43\nsp\MeV} \right)^{4}.
\end{equation}
Integrating equation (\ref{e.prethresholdrate}) over $\Sigma$ and using equation (\ref{e.Sigmaofchempot}), 
\begin{equation}\label{e.ecrateintegrated}
\int_{0}^{\chempot} \ecrate (\chempot',T)\,d\Sigma' \approx \left(\frac{4 \kB T}{\chempot} \right) \ecrate(\chempot,T) \Sigma
\end{equation}
for $\chempot \gg \kB T$.  Equation (\ref{e.ecrateintegrated}) shows that prethreshold electron captures occur in a narrow column depth range 
\begin{equation}\label{e.deltaSigmaoverSigma}
\frac{\Delta \Sigma}{\Sigma} \approx \frac{4 \kB T}{\chempot} = 0.050\left (\frac{\Teight}{5} \right ) \Stwelve^{-1/4},
\end{equation}
where $T = \Teight \times 10^{8}\nsp \K$.

Now consider electron captures in steady-state, such that electrons capture onto nuclei at the same rate as accretion advects the nuclei \citep[see also the discussion in][]{bildsten98}.  Equation (\ref{e.continuity}) becomes
\begin{equation}\label{e.steady-statecontinuity}
\sdot \pd{\ln X}{\Sigma} = - \ecrate.
\end{equation}
Integrating equation (\ref{e.steady-statecontinuity}) from $0$ to $Q$ and using equations (\ref{e.Sigmaofchempot}) and (\ref{e.ecrateintegrated}), we find that most electron captures occur prethreshold when
\begin{equation}\label{e.eccolumndepth}
\Stwelve > 0.027 \left (\frac{\Teight}{5} \right )^{-16/5} \left ( \frac{\sdot}{0.3 \sdotEdd}\right )^{4/5} \left (\frac{\ft}{10^{3} \nsp\unitstyle{s}} \right )^{4/5} ,
\end{equation}
where $\sdotEdd \approx 10^{5} \nsp\gram\usp\power{\cm}{-2}\usp\power{\unitstyle{s}}{-1}$ is the local accretion rate at which the accretion flux equals the Eddington flux\footnote{%
We define the Eddington luminosity to be $4\pi GMc/\kappa_{\mathrm{es}}$ in the frame of a distant observer. This is the largest luminosity observable by such an observer \citep[see][]{shapiro83}.  This differs from the definition used by \citet{Galloway2008Thermonuclear-t} by a factor of $[1-2GM/(Rc^{2})]^{-1/2} = 1.3$.
}.  Superbursts ignite at column depths $\Stwelve \sim 1$ and accretion rates $\sdot \approx 0.1$--$1 \nsp \sdotEdd$.  Equation (\ref{e.eccolumndepth}) shows that superallowed electron captures (for which $\ft \sim 10^{3}$--$10^{4}\nsp\unitstyle{s}$) occur prethreshold at superburst ignition depths.  

\subsection{Thermal Stability Analysis}\label{s.stabilityanalysis}

We now derive a one-zone model \citep[e.g.,][]{fujimoto81:_shell_x,Paczynski1983A-one-zone-mode,bildsten:thermonuclear} from the governing equations (\ref{e.transport}-\ref{e.continuity}) to determine the stability of prethreshold electron captures to thermal perturbations and thereby ascertain whether electron captures trigger superbursts.  
We consider only temperature perturbations and ignore the accretion-induced entropy advection through the bottom of the zone.  Therefore, we set $\partial/\partial t = 0$ in equation (\ref{e.continuity}) and approximate $Ds/Dt = \partial s/\partial t$ in equation (\ref{e.entropy}).  Perturbations occur at constant pressure since the scale height $\sim \Sigma/\rho \ll R$; therefore, we write $Tds = C_{P} dT$, where $C_P$ is the specific heat at constant pressure.  Equations (\ref{e.transport}-\ref{e.continuity}) become 
\begin{eqnarray}
F &=& \rho K \pd{T}{\Sigma}, \label{e.transport2}\\
C_P \pd{T}{t} &=& \estar X \ecrate + \pd{F}{\Sigma}, \label{e.entropy2}\\
\sdot \pd{X}{\Sigma} &=& -X \ecrate\label{e.continuity2}.
\end{eqnarray}

We simplify equations (\ref{e.transport2}-\ref{e.continuity2}) as follows.  We set $\rho$, $F$, and $X$ to be constant throughout the layer; specifically, we adopt step-like profiles for $F$ and $X$:
\begin{equation}
F(\Sigma,t) = F_0(t) \Theta(\Sigmaec-\Sigma),\label{e.FofSigma} \quad
X(\Sigma) = X_0 \Theta(\Sigmaec-\Sigma), \label{e.XofSigma}
\end{equation}
where $\Theta$ is the Heaviside step function, $\Sigmaec$ denotes the column depth at the bottom of the layer, and $F_0$ and $X_0$ denote the values at the top of the layer, where $\Sigma \ll \Sigmaec$.  We assume the ocean consists of a single ion species and set $X_0 = 1$.  Electron-ion scattering sets the ocean's thermal conductivity \citep{yakovlev80:_therm,itoh83,potekhin99:_trans}
\begin{equation}\label{e.K}
K \approx \frac{9.8\ee{18}}{ Z^{2/3} A^{1/3}} \left (\frac{\Teight}{5} \right )\left (\frac{\rho}{6\ee{8}\nsp \grampercc } \right )^{1/3} \mathrm{ergs~cm^{-1}~s^{-1}~K^{-1}},
\end{equation}
where we set the Coulomb logarithm $\Lambda_{\mathrm{ei}} = 1$, a value appropriate for a plasma at $\Teight \approx 5$ and $\rho \approx 6\ee{8}\nsp\grampercc$.  Since $K \propto T$, we rewrite equation (\ref{e.transport2}) as 
\begin{equation}\label{e.transport3}
F = \frac{\rho K}{2 T} \pd{T^{2}}{\Sigma}.
\end{equation}
Equations (\ref{e.FofSigma}) and (\ref{e.transport3}) then imply
\begin{equation}\label{e.TofSigma}
T^{2}(\Sigma,t) = T_0^{2} + \left [\Tec(t)^{2}-T_0^{2} \right ]\frac{\Sigma}{\Sigmaec}.
\end{equation}
Integrating equations (\ref{e.entropy2}-\ref{e.continuity2}) over $\Sigma$ and using equations (\ref{e.ecrateintegrated}), (\ref{e.deltaSigmaoverSigma}), (\ref{e.XofSigma}), (\ref{e.transport3}), and (\ref{e.TofSigma}), we find
\begin{eqnarray}
C_P \pd{\Tec}{t} &=& \frac{\Delta \Sigma}{\Sigmaec}\estar \ecrate- \frac{\rho K}{2 T}\left(\frac{\Tec^{2}-T_0^{2}}{\Sigmaec^{2}} \right ),\label{e.entropy3} \\
\ecrate &=& \frac{\sdot}{\Delta \Sigma} \label{e.continuity3}.
\end{eqnarray}
Equation (\ref{e.continuity3}) shows that electron capture occurs when the lifetime of a parent nucleus $1/\ecrate$ equals the time $\tacc \Delta \Sigma/\Sigma$ an advecting element spends within the capture region, where 
\begin{equation}\label{e.tacc}
\tacc \equiv \Sigma/\sdot
\end{equation}
is the accretion timescale.  Note that this differs from the usual assumption \citep{blaes90:_slowl,bildsten98:gravity-wave,bildsten98,ushomirsky.cutler.ea:deformations} that electron capture occurs when $1/\ecrate = \tacc$.

Finally, conducting a linear stability analysis on equation (\ref{e.entropy3}) and using equation (\ref{e.continuity3}), the thermal instability criterion is 
\begin{equation}\label{e.stabilitycriterion}
\nu \frac{\estar}{\tacc} > \frac{\rho K \Tec}{\Sigmaec^{2}},
\end{equation}
where the temperature-sensitivity of the height-integrated electron capture rate 
\begin{equation}\label{e.nu}
\nu \equiv  \pd{\ln (\ecrate \Delta \Sigma)}{\ln T} = 4 + \ln\left[8\ln2\frac{\tacc}{\ft}\frac{Q^{2}(\kB T)^{4}}{(\me c^{2})^{5}\chempot} \right ]
\end{equation}
from equations (\ref{e.prethresholdrate}) and (\ref{e.ecrateintegrated}).  Noting that electrons capture when $\chempot/Q \approx 1$, we write equation (\ref{e.nu}) as
\begin{equation}
\nu = 8.14+\ln\left [ \left (\frac{\Teight}{5} \right )^{4} \Stwelve^{5/4} \left( \frac{\sdot}{0.3 \sdotEdd}\right )^{-1} \left (\frac{\ft}{10^{3} \nsp\unitstyle{s}} \right )^{-1}     \right ].
\end{equation}
Using equations (\ref{e.tacc}) and (\ref{e.K}), the thermal instability criterion (\ref{e.stabilitycriterion}) becomes
\begin{equation}\label{e.stabilitycriterion2}
\left(\frac{\nu}{8.14}\right ) \Qstar > 20 \left (\frac{\Teight}{5} \right )^{2}\left( \frac{\sdot}{0.3 \sdotEdd}\right )^{-1} \left( \frac{A}{2Z} \right )^{2} \nsp \MeV,
\end{equation}
where $\Qstar$ is the energy released per electron capture.

We tested the accuracy of equation (\ref{e.stabilitycriterion}) using the suitably modified global linear stability analysis of \citet{cooper.narayan:theoretical}.  The minimum $\Qstar$ for instability derived from the global stability analysis differed from that of equation (\ref{e.stabilitycriterion2}) by less than $30\%$ for each of the 12 test cases.

Typically $\Qstar < Q$ because the daughter nucleus is generally more massive than the parent nucleus, although exceptions exist.  \citet{Gupta2006Heating-in-the-} find $\Qstar < 6.2\nsp\MeV$ for all electron captures that occur for $\chempot < 6\nsp\MeV$, or equivalently, $\Stwelve < 10$ (eq.~[\ref{e.Sigmaofchempot}]).  From equation (\ref{e.stabilitycriterion2}), it follows that  electron captures are thermally stable for the accretion rates and column depths at which superbursts occur.  Therefore, we conclude that electron captures do not trigger superbursts.

\section{Alternative Fusion Reactions}\label{s.light-fusion}

In this section, we examine whether light-element fusion reactions trigger superbursts.  Hydrogen has an electron capture threshold energy $Q=1.2933 \nsp \MeV$ and thus depletes at $\Stwelve \lesssim 2\ee{-2}$ (eq.~[\ref{e.Sigmaofchempot}]).  The helium abundance at $\igncol$ is less certain, and we discuss it below.  The paucity of stable isotopes of $Z=3\textrm{--}5$ nuclei leaves \carbon\ as the next reasonable alternative, which we address in \S \ref{s.carbon-sigma}. Finally, nuclei with $Z>6$ are unlikely candidates because the extra Coulomb repulsion causes the fusion rates to be significantly lower than that of  \carbon. 

Thus, other than \carbon\ + \carbon, $\alpha$ capture reactions such as $\carbon(\alpha,\gamma)\oxygen$ are the only plausible fusion reactions that might trigger superbursts. The conditions for these reactions to produce superbursts are similar to those for electron captures, namely, that (1) the reaction rate $\nucrate$ is sufficiently temperature dependent to produce unstable burning, and (2) $\alpha$ particles must survive to the inferred $\igncol$. Below, we show that the latter condition is not met; thus $\alpha$ particles deplete too quickly to trigger superbursts.

The condition for $\alpha$ particles to survive at a column depth $\Sigma$ is $Y / \sum_i \left(
\nucrate \right)_i > \Sigma/\sdot$ (see \S \ref{s.stabilityanalysis}), where $Y$ is the helium mass fraction and the sum is over all reactions that consume $\alpha$ particles. Using the triple-$\alpha$ reaction rate of \citet{fushiki87:_s} and setting $\rho = 6\ee{8} \nsp \grampercc$, $\Teight = 5$, and $Y=1$, we find $\nucrate = 2.2 \times 10^{6}~\mathrm{s}^{-1}$, which implies a lifetime of $4.6 \times 10^{-7}~\mathrm{s}$. The reaction rate $\nucrate \propto Y^{3}$, so the lifetime is much larger for smaller helium abundances. The accretion timescale $\tacc = 10^{7} [\Stwelve/(\sdot/\sdotEdd) ] \nsp \mathrm{s}$ (eq.~[\ref{e.tacc}]), indicating that $Y< 10^{-7}$ for helium to survive. At this low $Y$, the rise in temperature from consuming the helium via, e.g., $\carbon(\alpha,\gamma)$, is $\lesssim 10^{6}\nsp\K\ll T$ and hence insufficient to trigger a thermal instability.  

From the results of this section and \S \ref{s.ecZ}, we conclude that \carbon\ fusion triggers superbursts. 

\section{The \carbon\ + \carbon\ Reaction Rate}\label{s.carbon-sigma}

A possible solution to the superburst ignition problem is that the true \carbon\ + \carbon\ fusion rate is larger than assumed.  \carbon\ ignition at the inferred $\igncol$ requires a $\sim 10^{4}$ reaction rate enhancement for an ocean temperature of $4\ee{8}\nsp \K$ \citep{Cumming2005Long-Type-I-X-r} or a $\sim 10^{2}$ enhancement for $5\ee{8}\nsp \K$.  The two sources of uncertainty in the fusion rate are (1) plasma screening effects and (2) the nuclear cross-section $\sigma (E)$.  In the following subsections, we investigate whether either source could account for such a large increase in the fusion rate.

\subsection{Plasma Screening}\label{s.screening}

Superbursts ignite in a strongly-coupled Coulomb plasma. Two dimensionless parameters determine the plasma's state.  The first is the Coulomb coupling parameter 
\begin{equation}\label{e.Gamma}
\Gamma \equiv \frac{Z^{2}e^{2}}{a\kB T} = 6.0 \left (\frac{\Teight}{5} \right )^{-1} \Stwelve^{1/4} \left (\frac{Z}{6} \right)^{5/3},
\end{equation}
where $a=(3Z/4\pi n_e)^{1/3}$ is the ion-sphere radius, $n_e$ is the electron number density, and we used equation (\ref{e.approx-rho}), which assumes the gravitational acceleration $g = 2.43\ee{14} \nsp \cm \usp \power{\unitstyle{s}}{-2}$.  
For $\Gamma \ll 1$ Coulomb coupling is weak and the ions constitute a Maxwell-Boltzmann gas.  As 
$\Gamma$ increases, the ions gradually become a Coulomb liquid.  When $\Gamma > 175$ 
the ions crystallize \citep{potekhin.chabirer:eos_solid}.  Equation (\ref{e.Gamma}) implies that superbursts ignite in a Coulomb liquid.  The second dimensionless parameter,
\begin{equation}\label{e.zeta}
\zeta  \equiv \frac{3\Gamma}{\tau} \approx 0.17   \left (\frac{\Teight}{5} \right)^{-2/3} \Stwelve^{1/4} \left(\frac{2Z}{A} \right)^{1/3},
\end{equation}
is the ratio of the classical turning point to the ion separation, where 
\begin{equation}\label{e.tau}
\tau = \left(\frac{27 \pi^2 \mu Z^{4} e^{4}}{2  \kB T \hbar^{2}}\right)^{1/3} = \left(\frac{27 \pi^2 A\mb Z^{4} e^{4}}{4  \kB T \hbar^{2}}\right)^{1/3}
\end{equation}
and $\mu$ is the reduced mass of the reacting nuclei.  Specifically, $\zeta = \rtp/a$,
where $\rtp$ is the radius at which the Coulomb energy $Z^{2}e^{2}/\rtp$ equals the 
classical Gamow peak energy \citep{C83}
\begin{equation}\label{e.Gamowpeak1}
\Epk = \frac{\tau \kB T}{3}.
\end{equation}

Many-body interactions in a strongly-coupled Coulomb plasma modify the Coulomb potential between two reacting nuclei \citep[for a review, see][]{I93}.  From these many-body interactions, one derives an effective two-body potential
\begin{equation}\label{e.two-bodypotential}
V(r) = \frac{Z^{2}e^{2}}{r} - H(r),
\end{equation}
where $r$ is the distance between the reacting nuclei.  From $H(r)$, one derives the plasma screening enhancement to the reaction rate $\exp(\langle H(r)\rangle/\kB T$), where $\langle H(r) \rangle$ is a path-integral average of $H(r)$ \citep[e.g.][]{I93}.  One can expand the static mean-field potential $H(r)$ as a power series in $(r/a)^{2}$ \citep{W63}.  Neglecting quantum effects in $H(r)$, the leading order term $H(0)$ is a thermodynamic quantity; $H(0)$ equals the difference between the Coulomb (or excess) Helmholtz free energy before and after the reaction \citep{dewitt.graboske.ea:screening}.  

Monte Carlo simulations and hypernetted chain calculations of binary ionic mixtures \citep{Hansen1976Equation-of-sta,HTV77,CA90,ogata.iyetomi.ea:equations,R95,Rosenfeld1996Equation-of-sta,DSC96,DS03} suggest the excess free energy obeys the linear mixing rule to high accuracy in the regime $\Gamma > 1$ \citep{PCR09}.  Therefore, authors usually invoke the linear mixing rule when deriving the plasma screening enhancement to the reaction rate.  In this case, the total free energy of an ionic mixture 
\begin{equation}\label{e.Fex}
\frac{\Fex}{\kB T} =  \displaystyle\sum_{i} N_i \fex(\Gamma_i),
\end{equation}
where $N_i$ is the number of ions with charge $Z_i$, $\fex \equiv F^{\mathrm{ex, OCP}}/N\kB T$ is the well-determined reduced excess free energy per ion of a one-component plasma \citep[e.g.,][]{chabrier98,potekhin.chabirer:eos_solid}, and $\Gamma_i \propto Z_i^{5/3}$ is the Coulomb coupling parameter for species $i$ (eq.~[\ref{e.Gamma}]).  

From equation (\ref{e.Fex}), $H(0)/\kB T = 2\fex(\Gamma)-\fex(2^{5/3}\Gamma)$ \citep[][see also the Appendix]{J77}, where $\Gamma$ is that of the reacting ions; using the ion-sphere model result $\fex \approx -0.9 \Gamma$ \citep{Salpeter1954Electrons-Scree}, $H(0)/\kB T = 0.9(2^{5/3}-2)\Gamma \approx 1.0573\Gamma$, so the lowest-order screening enhancement to the \carbon\ + \carbon\ reaction rate
\begin{equation}\label{e.S}
\exp \left (\frac{H(0)}{\kB T} \right) = 5.5\ee{2} \exp \left[ \left (\frac{\Teight}{5} \right )^{-1} \Stwelve^{1/4} - 1 \right].
\end{equation}
Despite its simplicity, equation (\ref{e.S}) is adequate for most applications \citep{Gasques2005Nuclear-fusion-,Yakovlev2006Fusion-reaction,CDY07}.  We discuss corrections to the screening enhancement below.

\subsubsection{Deviation from Linear Mixing Rule}\label{s.LMRdeviations}

The excess free energy $\Fex$ of a multicomponent plasma exhibits small deviations from the linear mixing rule (eq.~[\ref{e.Fex}]).  In general, 
\begin{equation}\label{e.Fex2}
\frac{\Fex}{\kB T}  = \displaystyle\sum_{i} N_i \fex(\Gamma_i) +  N \Delta \fex,
\end{equation}
where $\Delta \fex \geq 0$ is a function of both the charges $Z_i$ and concentrations $x_i \equiv N_i/N$ of the ionic species \citep[e.g.,][]{DSC96}.  Using the hypernetted chain calculations of \citet{DSC96} and the ansatz $\Delta \fex \propto x_1x_2 (Z_2/Z_1)^{3/2}$ \citep{DS03}, we find
\begin{equation}\label{e.deltafex}
\Delta \fex = (0.0091 \ln \Gamma_1 + 0.018) x_1 x_2 \left(\frac{Z_2}{Z_1}\right )^{3/2}
\end{equation}
for a binary ionic mixture \citep[see also][]{PCR09}.  

To incorporate linear mixing rule deviations into $H(0)$ calculations, previous authors assumed a one-component plasma (consisting of \carbon\ ions in this case).  Fusion of two \carbon\ ions generates a compound nucleus \magnesium\ and thereby forms a binary ionic mixture.  One determines $H(0)$ by finding the difference in $\Fex$ before and after the reaction in the limit that the compound nucleus concentration $x_2 \rightarrow 0$ \citep{I93}.  Using this assumption and equations (\ref{e.Fex2}) and (\ref{e.deltafex}), we find the correction to $H(0)$: 
\begin{equation}\label{e.H0BIM}
\frac{\Delta H(0)}{\kB T} =  -0.026\ln \Gamma - 0.051,
\end{equation}
in good agreement with \citet{DSC96}; for a one-component plasma, linear mixing rule deviations reduce the plasma screening enhancement factor by $\sim 10\%$.  

However, the plasma at the superburst ignition depth is likely a mixture of \carbon\ and heavier ions with $Z \lesssim 46$ \citep[e.g.,][]{koike99,KHKF04,schatz.aprahamian.ea:endpoint,Schatz2003Nuclear-physics,woosley.heger.ea:models}.  Generalizing equation (\ref{e.deltafex}) for a multicomponent plasma with $Z_i < Z_j$ for $i<j$, we find \citep{ogata.iyetomi.ea:equations}
\begin{equation}\label{e.deltafex_multi}
\Delta \fex = \displaystyle\sum_{i<j}x_i x_j \Delta \fex_{ij}; \quad
 \Delta \fex_{ij} = (0.0091 \ln \Gamma_i + 0.018) \left(\frac{Z_j}{Z_i}\right )^{3/2}.
\end{equation}

To illustrate the effect spectator ions have on the screening enhancement, consider for simplicity a ternary ionic mixture of \carbon, \magnesium, and a representative spectator ion \iron.  Using equations (\ref{e.Fex2}) and (\ref{e.deltafex_multi}) and again taking the \magnesium\ concentration $x_2 \rightarrow 0$, so that $x_1 + x_3 = 1$, we find (see Appendix)
\begin{equation}\label{e.H0TIM}
\frac{\Delta H(0)}{\kB T} = - \Delta \fex_{12}+x_3 \left [ \Delta \fex_{12} + (1+x_3)\Delta \fex_{13} -\Delta \fex_{23} \right ].
\end{equation}
Note that equation (\ref{e.H0TIM}) reduces to (\ref{e.H0BIM}) in the limit $x_3 \rightarrow 0$, as it should.  Equation (\ref{e.H0TIM}) shows that, since the bracketed term is positive, heavy spectator ions {\it increase} the screening enhancement factor (i.e. $\partial \Delta H(0)/\partial x_3 > 0$).  For the fiducial \carbon\ mass fraction $0.2$ and \iron\ mass fraction $0.8$ \citep[such that $x_1 = 7/13$ and $x_3 = 6/13$;][]{Cumming2005Long-Type-I-X-r}, linear mixing rule deviations increase the plasma screening enhancement by $\approx 10 \%$.

\citet{PCR09} developed an analytic formula for $\delfex$ that is more accurate than equation (\ref{e.deltafex_multi}).  We derive the corresponding formula for $\Delta H(0)/\kB T$ in the Appendix.  In Figure \ref{f.deltaH0}, we plot $\Delta H(0)/\kB T$ as a function of the spectator ion number fraction $x_3$ using both equation (\ref{e.H0TIM}) and the expression derived from \citet{PCR09}.  Figure \ref{f.deltaH0} confirms that heavy spectator ions increase the plasma screening enhancement to the reaction rate, although the two expressions for $\Delta H(0)/\kB T$ differ quantitatively.

\begin{figure}
\epsscale{1.1}
\plotone{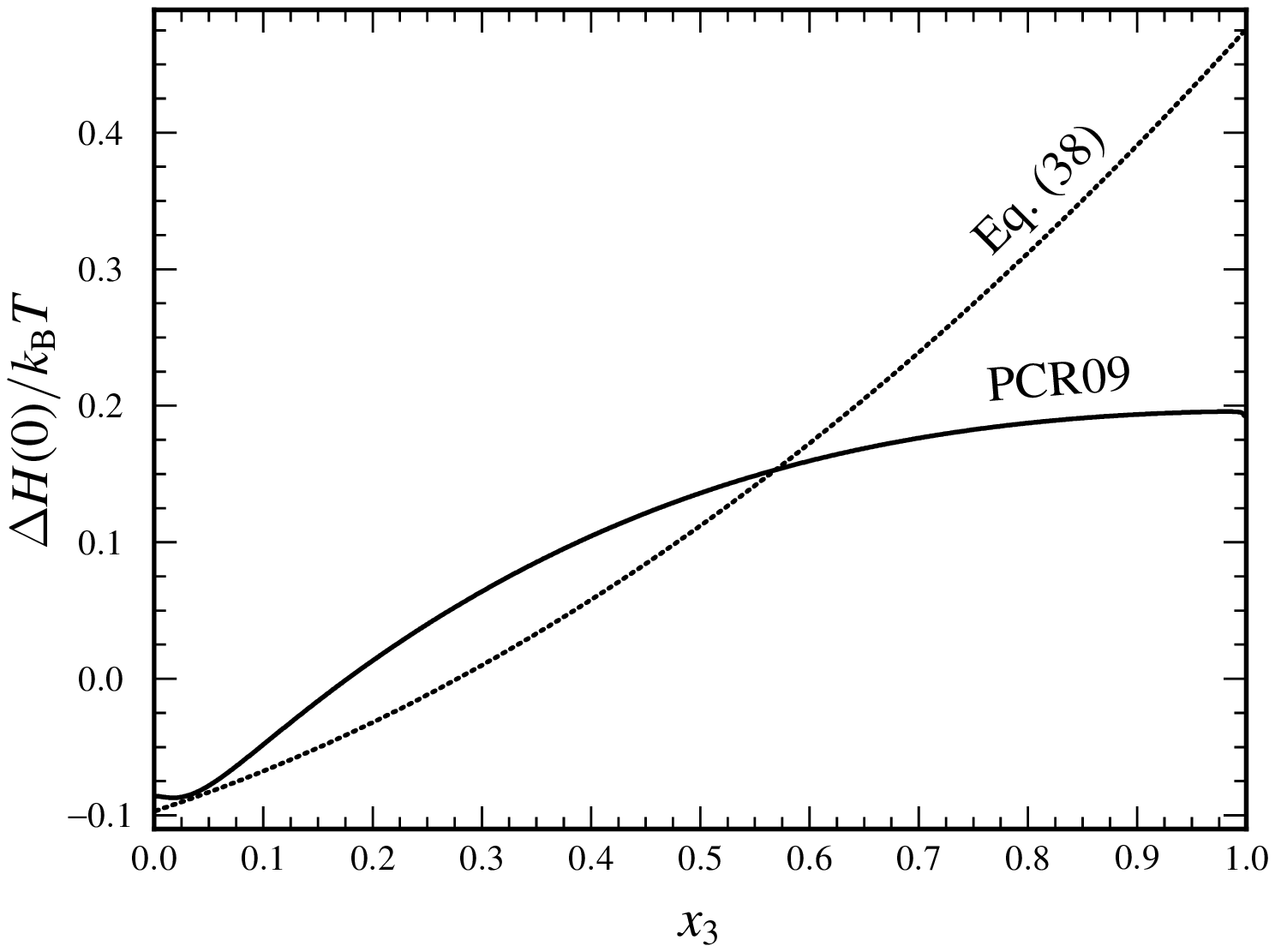}
\caption{Correction to the plasma screening enhancement factor due to linear mixing rule deviations as a function of the spectator ion number fraction $x_3$.  Considered is a ternary ionic mixture with $Z_1 = 6$, $Z_2 = 12$, $Z_3 = 26$, and $\Gamma_1 = 6$.  The number fraction of the product ion $x_2 = 0$, so $x_1 = 1-x_3$.  "PCR09" refers to the expression for $\Delta H(0)/\kB T$ derived from the results of \citet[][see Appendix]{PCR09}.
}
\label{f.deltaH0}
\end{figure}

\subsubsection{Corrections to $\langle H(r)\rangle$}

The next term in the expansion of $H(r)$ goes as $(r/a)^{2} \propto \zeta^{2}$; its contribution is small because $\zeta^{2} \ll 1$ (eq.~[\ref{e.zeta}]).  From \citet{J77}, we find
 \begin{equation}
\frac{\langle H(r)\rangle-H(0)}{\kB T} \approx -\frac{5}{32}\Gamma \zeta^{2} = -0.027 \left (\frac{\Teight}{5} \right )^{-7/3} \Stwelve^{3/4}.
 \end{equation}
This result agrees very well with more accurate calculations \citep{AJ78,OII91,Ogata1997Enhancement-of-}.  Higher order terms are even smaller; therefore, we conclude that corrections to $\langle H(r)\rangle$ are unimportant in calculating the plasma screening enhancement for \carbon\ ignition.

\subsubsection{Electron Screening Corrections}

In the above analysis, we tacitly assumed a uniform electron density.  Although highly degenerate, electrons nonetheless slightly concentrate around positively charged ions.  Electron polarization mitigates the Coulomb repulsion between ions relative to an unpolarized configuration.  This has two counteracting effects: It (1) lowers the Coulomb repulsion between the two reacting ions, which increases the reaction rate, and (2) attenuates the many-body Coulomb interactions and thereby $H(r)$, which decreases the reaction rate.  The Yukawa potential $Z^{2}e^{2}/r \exp{(-r/r_{\mathrm{TF}})}$ describes the two-body potential, where $r_{\mathrm{TF}}$ is the Thomas-Fermi screening length.  For relativistic, degenerate electrons, $r_{\mathrm{TF}}/a = 3.0 (Z/6)^{-1/3}$ \citep[e.g.,][]{HPY07}, so electron screening is weak; from the results of \citet{SC98}, electron screening changes the reaction rate by $\lesssim 1\%$.
\linebreak

Corrections to the lowest-order plasma screening enhancement (eq.~[\ref{e.S}]) change the \carbon\ + \carbon\ reaction rate by a factor $< 2$.  Therefore, we conclude that uncertainties in the plasma screening enhancement are too small to explain the discrepancy between superburst observations and theoretical model results.

\subsection{The Nuclear Cross-Section}\label{s.sfactor}

Although the plasma screening enhancement to the \carbon\ + \carbon\ reaction rate is well-determined for superburst conditions, the nuclear cross-section $\sigma(E)$ is not.  Many groups have measured $\sigma(E)$ at various center-of-mass energies $E$ down to $\approx 2.1\nsp\MeV$ \citep{PWZ69,MS72,MS73,SW74,HC77,KLRW77,KLR80,Eetal80,TFGDV80,BKRT81,DCL82,SDKX82,Retal03,B-Petal04,B-Petal06,Aetal06,Spillane2007C12C12-Fusion-R}.
However, the energy range of interest is centered at the classical Gamow peak energy (cf. eq.~[\ref{e.tau}--\ref{e.Gamowpeak1}])
\begin{equation}
\Epk = 1.5\left (\frac{\Teight}{5} \right )^{2/3 }\nsp \MeV
\end{equation}
and has a full width
\begin{equation}
\Delta \Epk = 4 \left(\frac{\Epk \kB T}{3} \right )^{1/2} = 0.59
\left(\frac{\Teight}{5} \right )^{5/6}\nsp \MeV.
\end{equation}
Thus, $\sigma(E)$ in the astrophysically relevant energy range is experimentally unknown.

This situation is common in nuclear astrophysics: to determine the astrophysical reaction rate, one either extrapolates the experimental data to lower 
energies or calculates the rate theoretically \citep[e.g.,][]{caughlan88:_therm,Gasques2005Nuclear-fusion-}. In doing so, one tacitly assumes the astrophysical rate has the nonresonant form, i.e., no prominent resonances exist in the compound nucleus within the relevant energy range\footnote{This statement is not strictly true for heavy ion fusion reactions such as \carbon\ + \carbon.  The compound nucleus \magnesium\ has numerous quasi-stationary states at excitation energies near $15 \nsp \MeV$ above the ground state \citep{E90,F07}, where $\Epk$ lies; thus all reactions are resonant.  However, when the mean level spacing of quasi-stationary states $D \lesssim \kB T$, one computes an average cross-section over all resonances, and the reaction rate assumes the nonresonant form \citep[e.g.][]{Cameron1959Carbon-Thermonu,FH64}}.  However, several groups have detected strong resonances in the  \carbon\ + \carbon\ system at energies below the Coulomb barrier.  Resonances exist throughout the entire energy range probed so far, and the spacing between adjacent resonances\footnote{The average level spacing of the detected resonances is much greater than that of the quasi-stationary states in the compound nucleus \magnesium.  Thus the observed resonances are not ordinary compound nuclear states.  As \citet{ABK60} first suggested, the resonances are probably quasi-molecular doorway states in the \carbon\ + \carbon\ system \citep[e.g.,][]{BW97}.} is $\approx 0.3 \nsp \MeV$.  Therefore, a resonance probably exists near $\Epk$ \citep{BKA60,ABK60,GTDFV77,KEWB79,Eetal80,TFGDV80,Spillane2007C12C12-Fusion-R}.  Indeed, \citet{MV72} and \citet{PBA06} predict a resonance exists in the \carbon\ + \carbon\ system with energy $\ER \approx 1.5 \nsp \MeV$.  If the resonance is strong, the thermally averaged reaction rate $\sigmav$ would be much larger than assumed.  

To illustrate the effect a strong resonance within the Gamow window would have on the \carbon\ + \carbon\ reaction rate, we follow the prediction of \citet{PBA06} and assume the existence of a single, narrow resonance with $\ER = 1.5 \nsp \MeV$.  Then 
\begin{eqnarray}
\sigmav &=& \sigmav_{\mathrm{NR}} + \sigmav_{\mathrm{R}}, \\ \nonumber \\
\sigmav_{\mathrm{R}} &=& \left (\frac{\pi}{3 \mb \kB T}\right )^{3/2}  \hbar^{2} \og \exp \left(-\frac{\ER}{\kB T} \right ), 
\end{eqnarray}
where $\sigmav_{\mathrm{NR}}$ is the nonresonant contribution to the total reaction rate as given in, e.g., \citet[][]{caughlan88:_therm}, $\sigmav_{\mathrm{R}}$ is the resonant contribution, 
\begin{equation}
\og= 2(2J + 1) \frac{\GammaC(\GammaR-\GammaC)}{\GammaR} \approx 2(2J+1)\GammaC
\end{equation}
 is the resonance strength, $J$ is the total angular momentum of the resonance, $\GammaC$ is the entrance channel width, and $\GammaR \gg \GammaC$ is the resonance width.

The resonant contribution $\sigmav_{\mathrm{R}} \propto \og$.  Using the Breit-Wigner single resonance formula,
\begin{equation}\label{e.BWresonance}
\sigma (E) = \frac{\pi \hbar^{2}}{12 E \mb} \frac{\og \GammaR}{(E-\ER)^{2} + (\GammaR/2)^{2}} 
\end{equation}
\citep[e.g.,][]{C83}.  Evaluating equation (\ref{e.BWresonance}) at $E = \ER$, 
\begin{equation}
\og = 3.4\ee{-8} \left (\frac{\GammaR}{100\nsp \keV} \right) \left (\frac{\sigma(\ER )}{10^{-13}\nsp \mathrm{barn}} \right ) \left (\frac{\ER}{1.5\nsp \MeV} \right ) \nsp \eV,
\end{equation}
where we normalize the resonance width $\GammaR$ to that typical of known resonances \citep[see, e.g., Table IV of][]{Aetal06} and the cross-section at resonance $\sigma (\ER)$ to the approximate value \citet{PBA06} predict.  For this work, we adopt $\og = 3.4\ee{-8}\nsp \eV$ as the fiducial resonance strength.  

To determine an upper limit for $\og$, we demand that the resonance's contribution to the astrophysical $S$-factor at a given energy $E$, $\SR$, be less than the experimentally measured value $\Sexp$ for all $E \gtrsim 2.1 \nsp \MeV$, the lowest energy probed at the time of this writing.  The $S$-factor 
for \carbon\ + \carbon\ is
\begin{equation}\label{e.Sfactor}
S(E) = \sigma(E) E \exp \left [87.21 \left (\frac{E}{\MeV} \right )^{-1/2} + 0.46 \left (\frac{E}{\MeV} \right )\right ]
\end{equation}
\citep{PWZ69,C83}.  From equations (\ref{e.BWresonance}) and (\ref{e.Sfactor}), we write 
\begin{equation}\label{e.SR}
\SR = S(\ER) \frac{(\GammaR/2)^{2}}{(E-\ER)^{2} + (\GammaR/2)^{2}}.
\end{equation}
Using equations (\ref{e.BWresonance}), (\ref{e.Sfactor}), and (\ref{e.SR}), demanding that $\SR < \Sexp$, and noting that $(E-\ER)^{2} \gg (\GammaR/2)^{2}$, we find
\begin{eqnarray}\label{e.ogmax}
\og &<& 5.5\ee{-8} \left (\frac{\GammaR}{100\nsp \keV} \right)^{-1}\nonumber\\
&&\times\left [ \left (\frac{E-\ER}{\MeV} \right )^{2} \left (\frac{\Sexp}{10^{16} \nsp \MeV \usp \mathrm{barn}} \right ) \right ] \nsp \eV
\end{eqnarray}
for a resonance at $\ER = 1.5\nsp  \MeV$.  Equation (\ref{e.ogmax}) must be satisfied for all $E$.  According to the experimental data, the minimum value of the bracketed term is $\approx 1$ \citep[see, e.g., Fig.~4 of][]{Spillane2007C12C12-Fusion-R}, so 
\begin{equation}\label{e.ogmax2}
\og < 5.5\ee{-8} \left (\frac{\GammaR}{100\nsp \keV} \right)^{-1} \nsp \eV.
\end{equation}
If $\GammaR \approx 100 \nsp \keV$, then our fiducial strength $\og = 3.4\ee{-8}\nsp \eV$ is comparable to the maximum possible strength.  $\GammaR$ may be much smaller, however; the resonance at $2.14 \nsp \MeV$, the lowest-energy resonance known as of this writing, has a width $\GammaR < 12 \nsp \keV$ \citep{Spillane2007C12C12-Fusion-R}.  Therefore, we set $\og = 3.4\ee{-7}\nsp \eV$, which is ten times larger than our fiducial rate, as a reasonable upper limit.  

\begin{figure}
\epsscale{1.1}
\plotone{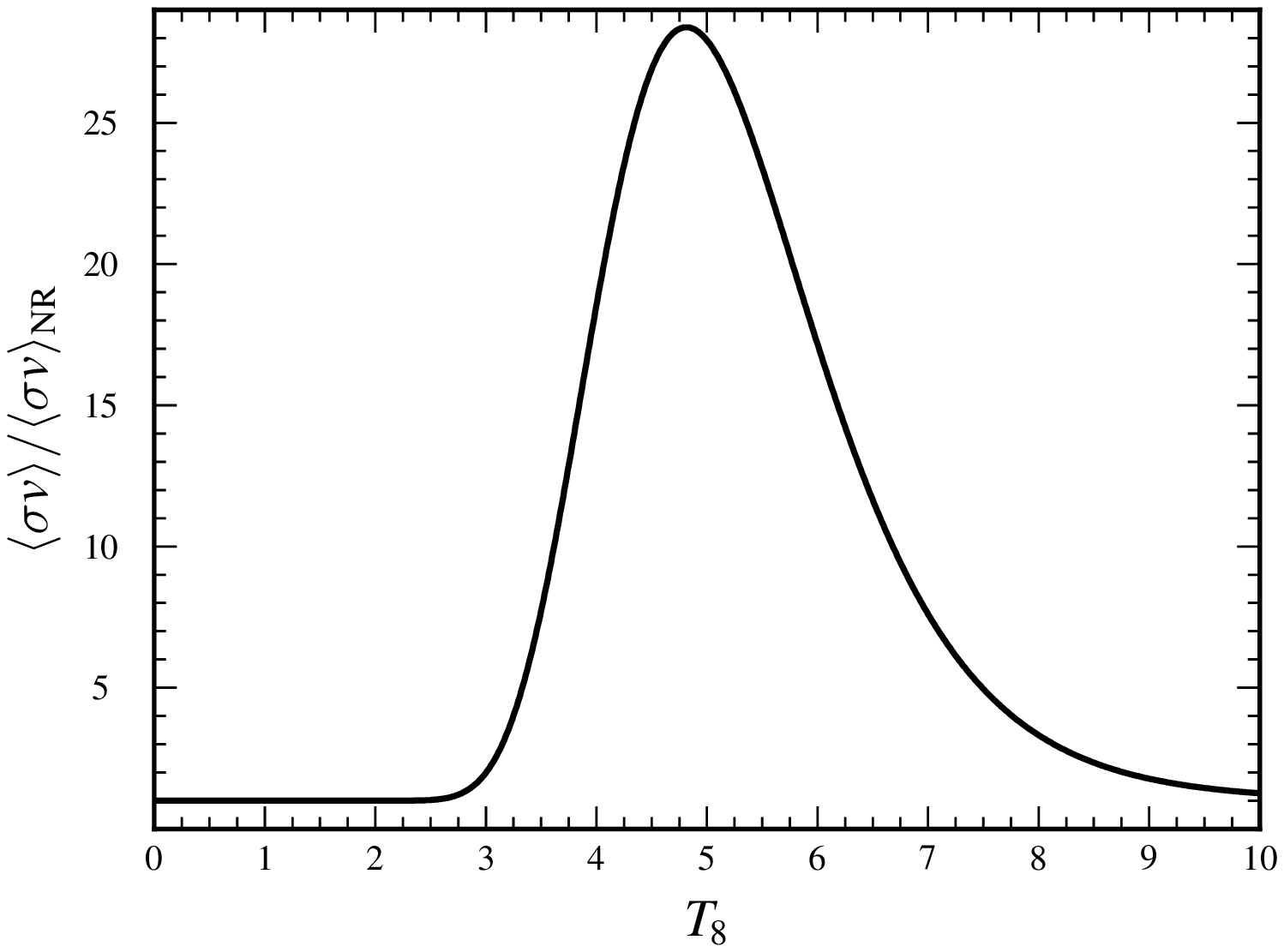}
\caption{Ratio of the total thermally averaged reaction rate $\sigmav = \sigmav_{\mathrm{NR}} + \sigmav_{\mathrm{R}}$ to the nonresonant contribution $\sigmav_{\mathrm{NR}}$ for a hypothetical $1.5\nsp \MeV$ resonance with strength $\og = 3.4\ee{-8}\nsp \eV$, the fiducial value, as a function of temperature $T = \Teight \times 10^{8}\nsp \K$.  The resonance increases $\sigmav$ by a factor $\gtrsim 25$ near $\Teight \approx 5$.  
}
\label{f.sigmav}
\end{figure}

Figure \ref{f.sigmav} shows the effect a $1.5 \nsp \MeV$ resonance has on the reaction rate $\sigmav$.  For the fiducial $\og$ value, the resonance increases $\sigmav$ by a factor $\gtrsim 25$ at temperatures relevant to superbursts; for the $\og$ upper limit, the resonance increases $\sigmav$ by a factor $\gtrsim 250$.  These increases are of the order required to reconcile the observationally inferred $\igncol$ with that calculated from theoretical models for a specific range of assumed crust thermal conductivities and core neutrino emissivities. In the following section, we compute the superburst $\igncol$ with the effect of this resonance.

\section{Effects of a Resonance on Superburst Ignition}\label{s.ignition-column}

We use the global linear stability analysis of \citet{cooper.narayan:theoretical} to determine the effect 
a strong resonance in the \carbon\ + \carbon\ system would have on the superburst ignition depth $\igncol$.  We assume steady spherical accretion onto a neutron star of mass $M = 1.4 \nsp M_{\odot}$ and radius $R = 10 \nsp \mathrm{km}$.  The accreted matter composition is that of the Sun: The hydrogen mass fraction $X = 0.7$, helium mass fraction $Y=0.28$, and heavy-element mass fraction $Z=0.02$.  Furthermore, we follow \citet{Cumming2005Long-Type-I-X-r} and assume the \carbon\ mass fraction $X_{\mathrm{C}} = 0.2$ at the base of the accreted layer.  

We make the following two modifications to the model of \citet{cooper.narayan:theoretical}. (1) \citet{cooper.narayan:theoretical} followed \citet{brown:nuclear} and assumed the energy generated by electron captures, neutron emissions, and pycnonuclear reactions in the crust was distributed uniformly between $\Stwelve = 6\ee{3}$ and $2 \ee{5}$.  We now follow \citet{Haensel2008Models-of-crust} and distribute the energy according to their Table A.3.  (2) Plasma screening reduces the entrance channel width $\GammaC$.  Therefore, the plasma screening enhancement for the resonant contribution to the reaction rate includes a correction factor that reduces the overall enhancement \citep{salpeter69:_nuclear,M77}, although the reduction is only a few percent for the conditions relevant for superbursts \citep[see, e.g., Fig.~1 of][]{CLL02}.  We now use the formalism of \citet{ITWN03} for the plasma enhancement factors of both the resonant and nonresonant contributions.

The \carbon\ + \carbon\ reaction rate, accretion rate $\sdot$, and ocean temperature profile together determine $\igncol$.  The temperature profile is a strong function of the crust's thermal conductivity and core's neutrino emissivity, both of which are poorly constrained.  We parametrize these uncertainties by implementing two conductivity and three core neutrino emissivity prescriptions that likely bracket their true values in accreting neutron stars.  The thermal conductivity is a decreasing function of the impurity parameter $\Qimp = \langle Z^{2} \rangle - \langle Z \rangle^{2}$ \nocite{DG09} \citep[][see also Daligault \& Gupta 2009]{itoh93}.   \citet{schatz99} found $\Qimp \sim 100$ from steady-state nucleosynthesis calculations, although subsequent calculations suggest $\Qimp$ should be smaller \citep{schatz.bildsten.ea:photodisintegration-triggered,woosley.heger.ea:models,KHKF04,Horowitz2007Phase-Separatio,Horowitz2008Thermal-conduct}. In addition, fits to the quiescent light curves of \source{KS}{1731}{-260} \citep{Shternin2007Neutron-star-co,BC09} and \source{MXB}{1659}{-29} \citep{BC09} require that $\Qimp \sim 1$. Since both observations and molecular dynamics simulations imply the crust forms an ordered lattice, 
we adopt  $\Qimp = 3$ and $100$ as the two bracketing values. The core neutrino emissivity, and thereby the core cooling rate, depends on the unknown ultradense matter equation of state \citep[for reviews, see][]{yakovlev.pethick:neutron,PGW06}.  We consider one ``fast'' cooling model for which the pion condensate process dominates and two "slow" cooling models for which either the modified Urca or nucleon-nucleon bremsstrahlung process dominates \citep[see, e.g., Table 1 of ][]{PGW06}; these roughly correspond to cases ``A,'' ``B,'' and ``D'' of \citet[][see their Table 2]{Cumming2005Long-Type-I-X-r}.  The respective core temperatures for these models are approximately $3\ee{7} \nsp \K$, $3\ee{8} \nsp \K$, and $6\ee{8} \nsp \K$.

\begin{figure}
\epsscale{1.1}
\plotone{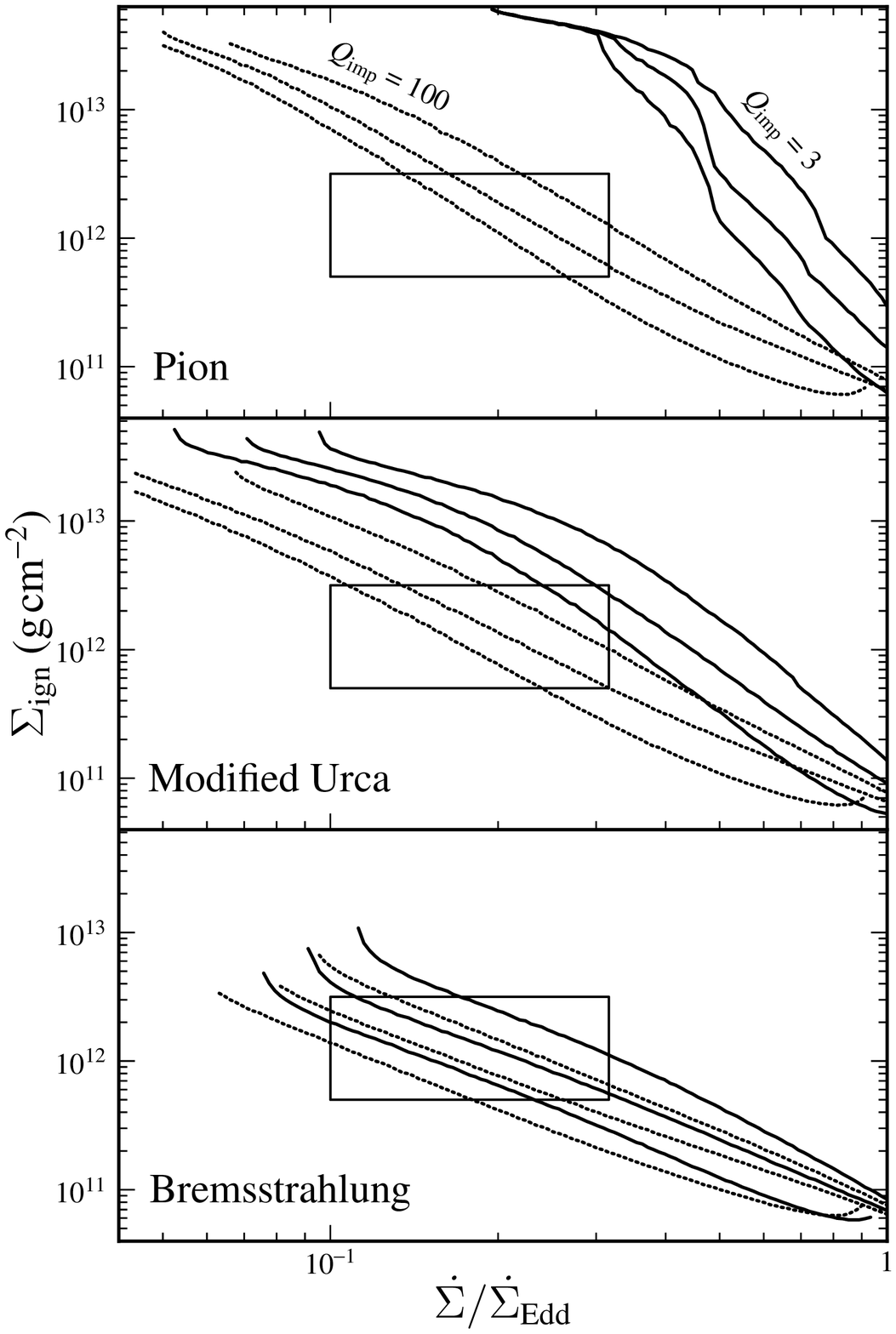}
\caption{Superburst ignition column depth $\igncol$ as a function of the Eddington-scaled accretion rate $\sdot/\sdotEdd$ for various model parameters.  Solid (dashed) lines show results for models with impurity parameter $\Qimp = 3$ $(100)$.  "Pion," "Modified Urca," and "Bremsstrahlung" refer to the core's dominant neutrino emission mechanism.  For a given $\Qimp$ and neutrino emission mechanism, the three lines show results for a \carbon\ + \carbon\ reaction rate with no resonances (the standard rate), a hypothetical $1.5 \nsp \MeV$ resonance with the fiducial strength $\og = 3.4\ee{-8}\nsp \eV$, and a hypothetical $1.5 \nsp \MeV$ resonance with an approximate maximum strength $\og = 3.4\ee{-7}\nsp \eV$, from top to bottom.  The boxes show the inferred $\igncol$ and $\sdot$ ranges for the majority of observed superbursts.  A $1.5 \nsp \MeV$ resonance lowers $\igncol$ by a factor $\approx 2$ and $\approx 4$ for the fiducial and maximum $\og$ values, respectively.
}
\label{f.igncol}
\end{figure}

Figure \ref{f.igncol} shows the superburst ignition column depth $\igncol$ as a function of $\sdot/\sdotEdd$ for various neutron star models\footnote{The critical $\sdot$ below which \carbon\ burns stably calculated in our global stability analysis is lower than that calculated in the one-zone model of \citet[][compare to their Fig.\ 15]{Cumming2005Long-Type-I-X-r}.  The reason is simple: Following \citet{cumming.bildsten:carbon}, they demand that the characteristic lifetime of a \carbon\ ion $X_{\mathrm{C}}/\nucrate > \tacc$.  However, $\nucrate$ depends exponentially on the density due to plasma screening (\S \ref{s.screening}), so \carbon\ burning occurs in a narrow column depth range, much like the electron captures discussed in \S \ref{s.stabilityanalysis}.  Thus, the proper criterion is  $X_{\mathrm{C}}/\nucrate > \tacc \Delta \Sigma/ \Sigma$, where $\Delta \Sigma /\Sigma$ is similar to the expression given in equation (\ref{e.deltaSigmaoverSigma}).  This proper criterion gives a lower critical $\sdot$, in accord with our results.}.  A $1.5 \nsp \MeV$ resonance in the \carbon\ + \carbon\ system lowers $\igncol$ by a factor $\approx 2$ and $\approx 4$ for the fiducial and maximum $\og$ values, respectively; the lowered $\igncol$ values are in accord with the observationally inferred values for a range of realistic neutron star model parameters.  Therefore, we conclude that (1) a strong resonance may exist at an energy $\approx 1.5\nsp \MeV$ above the \carbon\ + \carbon\ ground state, and (2) if such a resonance exists, it will mitigate the discrepancy between observationally inferred superburst ignition depths and those calculated from theoretical models.

For the low-mass X-ray transient \lmxbtrans, which exhibited a superburst, the thermal quiescent luminosity constrains the core temperature to be $\approx 2.5\ee{8}\nsp\K$. Fits to the superburst light curve find an ignition column $\igncol = (1.5\textrm{--}4.1)\ee{12}\nsp\columnunit$.  A resonance at $1.5\nsp\MeV$ could make the ignition temperature over this range as low as $(4.1\textrm{--}4.8)\ee{8}\nsp\K$, which is marginally consistent with the calculated crust temperature at the time of the superburst \citep{Keek2007First-Superburs}.

For the transient \ks, the timescale for the effective temperature to decrease implies $\Qimp\lesssim 1$, and the lowest observed effective temperature implies that the core temperature is $\lesssim 10^{8}\nsp\K$ \citep{Shternin2007Neutron-star-co,BC09}. Under these conditions, the temperature at $\Sigma\approx10^{12}\nsp\columnunit$ is unlikely to be $> 3\ee{8}\nsp\K$ and therefore too cold to match the inferred ignition depth, even if the proposed resonance exists. Recent theoretical calculations of nuclear reactions in the neutron star crust suggest that the pycnonuclear fusion of neutron-rich, low-$Z$ ions such as \oxygen[24] \citep{Horowitz2007Fusion-of-neutr} or reactions triggered by $\beta$-delayed neutron emissions \citep{GKM08} may provide a strong source of heating for the neutron star outer crust. Indeed, fits to the quiescent light curves of \ks\ and \mxb\ suggest that the heating in the outer crust is larger than can be accounted for from electron captures \citep{BC09}. Although a survey of neutron star models with this additional heating is outside the scope of this paper, we note that a strong resonance does alleviate the discrepancy in $\igncol$ even if it does  not entirely resolve it.

\section{Summary and Discussion}\label{s.discussion}

In this work, we reexamined the superburst trigger mechanism to address the discrepancy between observationally inferred superburst ignition column depths $\igncol$ and those calculated in theoretical models.  Motivated by the suggestion of \citet{kuulkers.ea:ks1731-superburst} and the similarity between inferred ignition column depths from different sources, we first explored the viability of thermally unstable electron captures as the trigger mechanism in \S \ref{s.ecZ}.  We found that electron captures are always thermally stable in accreting neutron star oceans; thus electron captures do not trigger superbursts.  We then investigated the viability of nuclear fusion reactions other than \carbon\ + \carbon.  Accretion-induced nuclear reactions deplete ions with $Z < 6$ at column depths $\Sigma \ll \igncol$, whereas ions with $Z > 6$ fuse at $\Sigma \gg \igncol$.  We therefore confirmed the proposal \citep{cumming.bildsten:carbon,strohmayer.brown:remarkable} that \carbon\ + \carbon\ triggers superbursts.  

We then examined the \carbon\ + \carbon\ fusion rate in \S \ref{s.carbon-sigma}, noting that superburst model results would be in accord with observations if the true fusion rate were greater than the standard rate by a factor $\gtrsim 10^{2}$.  Two factors determine the fusion rate: plasma screening effects and the nuclear cross-section $\sigma (E)$.  Uncertainties in, and corrections to, the plasma screening enhancement to the reaction rate alter the usual enhancement by a factor $< 2$ and thus cannot resolve the discrepancy between superburst observations and theoretical models.  However, uncertainties in $\sigma (E)$ are much larger; indeed, $\sigma (E)$ is experimentally unknown at astrophysically relevant energies.  We find that a strong resonance in the \carbon\ + \carbon\ system at an energy near $1.5 \nsp \MeV$ could increase the fusion rate by a few orders of magnitude at the temperatures relevant to superbursts.  Both theoretical optical potential models and extrapolations of existing experimental data suggest a resonance exists at an energy near $1.5 \nsp \MeV$.  If this is true and the resonance strength $\og$ is sufficiently large, it could eliminate the discrepancy between observationally inferred superburst ignition column depths and theoretical model results (see Figure \ref{f.igncol}).  

In \S \ref{s.introduction} we outlined three fundamental problems that exist with superburst ignition.  We address these problems below in the context of our results.

(1) The results of all previous superburst models imply that ocean temperatures are too low for \carbon\ ignition at the inferred $\igncol \approx 10^{12}\nsp \columnunit$. A strong resonance near $1.5 \nsp \MeV$ in the \carbon\ + \carbon\ system would decrease the temperature required for ignition at $\igncol \approx 10^{12}\nsp \columnunit$ from $\approx 6\ee{8} \nsp \K$ to $\approx 5\ee{8} \nsp \K$.  

(2) Heavy-ion fusion hindrance would imply that the standard $S$-factor overestimates the true \carbon\ + \carbon\ fusion rate.  This existence of such hindrance is currently speculative for both \carbon\ + \carbon\ in particular \citep{JRBJ07} and exothermic fusion reactions in general \citep{Jetal08,Setal08}.  Furthermore, the effect heavy-ion fusion hindrance has on resonant reactions is unknown.  Therefore, it is unclear whether heavy-ion fusion hindrance poses a problem for superburst ignition.

(3) The \carbon\ yield from nucleosynthesis models is often lower than that required for a thermal instability.  A strong resonance would reduce the minimum \carbon\ abundance required for a superburst, but by only a small amount.  Thus, this problem would be attenuated but not resolved.  

Our result has implications for $\carbon+\carbon$ reactions in other contexts, namely Type Ia supernovae and massive stellar evolution. In addition, we have also presented a general prescription for understanding the screening enhancement factor in a multicomponent plasma. We briefly describe each of these topics before concluding with an outlook on future measurements.

\subsection{Implications for Type Ia Supernovae and Massive Stellar Evolution}\label{s.implication}

\begin{figure}
\epsscale{1.1}
\plotone{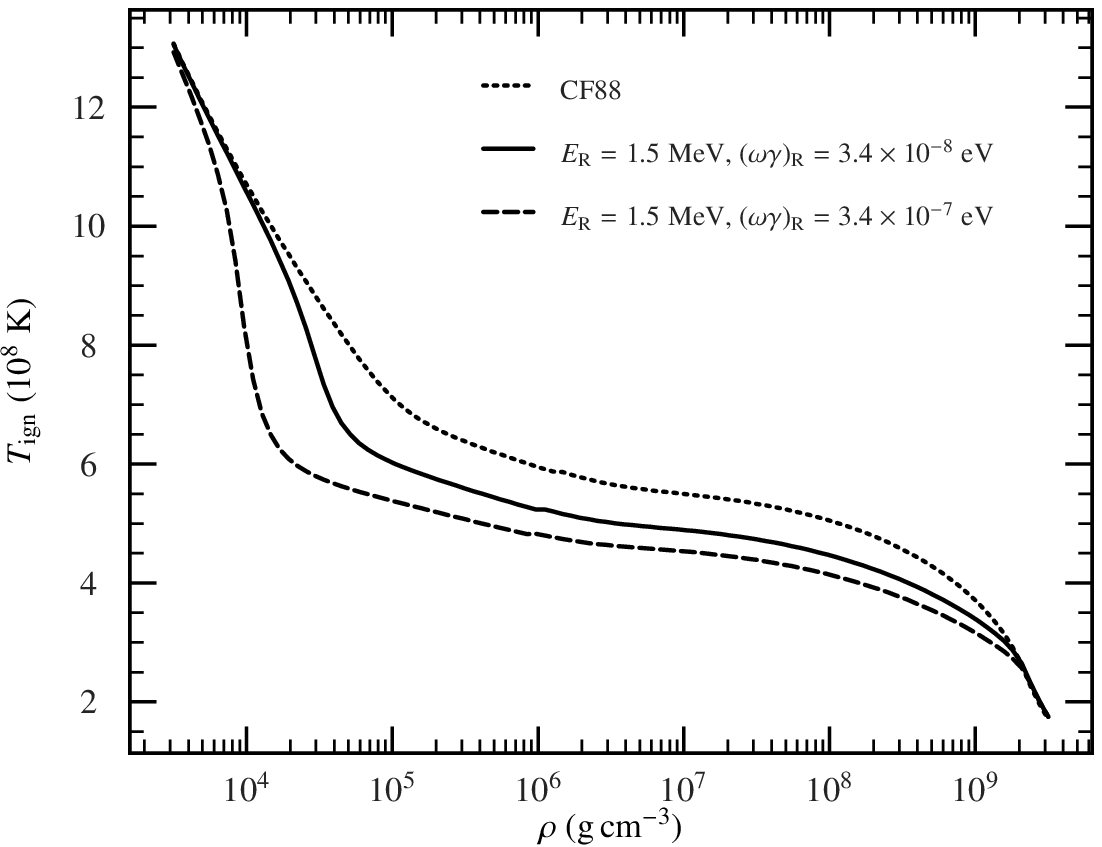}
\caption{Locus in the temperature-density plane where $\epsnuc = \epsnu$, which defines the ignition of \carbon\ for stellar burning in massive stars and for thermonuclear (Type Ia) supernovae. The composition is \carbon-\oxygen\ with $X_{\mathrm{C}}=0.5$. The three curves, from top to bottom, show $\epsnuc$ computed with the standard rate \protect\citep[\emph{dotted line}]{caughlan88:_therm}, with a resonance at our fiducial strength, $\og = 3.4\ee{8}\nsp\eV$ (\emph{solid line}), and with a resonance at our maximum strength, $\og = 3.4\ee{-7}\nsp\eV$ (\emph{dashed line}).
\label{f.ignition}}
\end{figure}

The fusion of \carbon\ is an important stage in the post--main-sequence evolution of a massive star, and it is the reaction that ignites a white dwarf and triggers a thermonuclear (Type Ia) supernova. In both systems, the competition between heating from the $\carbon+\carbon$ reaction and cooling from neutrino emissions determines ignition.  To explore the implications of a resonance in the reaction cross-section on these phenomena, we construct ignition curves (Fig.~\ref{f.ignition}), defined as $\epsnuc(\rho,T)=\epsnu(\rho,T)$, for a \carbon-\oxygen\ plasma with $X_{\mathrm{C}} = 0.5$. We compute the neutrino emissivity for the pair, photo, plasma, and bremsstrahlung processes using analytical fitting formulae \citep{itoh96:_neutr}. For $\epsnuc$, we use the effective reaction $\Qstar$-value of $9.0\nsp\MeV$ \citep{Chamulak2007The-Reduction-o}, which includes heating from both the $p$- and $\alpha$- branches and subsequent reactions; the ignition curve is insensitive to the choice of $\Qstar$.  Three curves are plotted in Fig.~\ref{f.ignition} for different choices of the $\carbon+\carbon$ rate: the standard nonresonant rate \citep[\emph{dotted line}]{caughlan88:_therm}, a resonance at $\ER=1.5\nsp\MeV$ with our fiducial strength $\og=3.4\ee{-8}\nsp\eV$ (\emph{solid line}), and a resonance at $\ER=1.5\nsp\MeV$ with our maximum strength $\og=3.4\ee{-7}\nsp\eV$ (\emph{dashed line}).

As is evident from Figure \ref{f.ignition}, the effect of a resonance at $\ER=1.5\nsp\MeV$ is minimal for the ignition of Type Ia supernovae, which are thought to ignite at central densities $>10^{9}\nsp\grampercc$ \citep[see, e.~g.,][]{Woosley2004Carbon-Ignition}. It is interesting to speculate that a resonance at a lower energy might shift the ignition curve to lower densities. This would reduce the \emph{in situ} neutronization of the nickel-peak material synthesized in the explosive burning and the neutronization during the pre-explosive convective burning \citep{Piro2007Neutronization-,Chamulak2007The-Reduction-o}; moreover, numerical simulations \citep{Ropke2005Type-Ia-superno} find that a lower central density reduces the growth of the turbulent flame velocity (because the lower gravitational acceleration decreases the growth rate of the Rayleigh-Taylor instability), which leads to a less vigorous explosion and a decreased production of iron-peak elements. Although lower densities are necessary to avoid overproduction of neutron-rich isotopes such as \ironfiftyfour\ and \nickelfiftyeight\ \citep{Woosley1997Neutron-rich-Nu,iwamoto.brachwitz.ea:nucleosynthesis}, this may be ameliorated by improved electron capture rates onto $pf$-shell nuclei \citep{martnez-pinedo.langanke.ea:competition,brachwitz.dean.ea:electron_captures}.  Moreover, there is observational evidence that the majority of supernovae do undergo electron-captures in the innermost $\approx 0.2\nsp\Msun$ of ejecta \citep{Mazzali2007A-Common-Explos}.
Given the uncertainties in modeling the progenitor evolution, flame ignition, and explosion, and in the dependence of the ignition density on the accretion history of the white dwarf \citep[see][for a recent discussion]{Lesaffre2006The-C-flash-and}, we do not think it possible to constrain the existence of such a resonance from observations at this time, but future modeling efforts should clearly allow for this possibility.

Intriguingly, the largest effect is at $\rho\lesssim 10^{5}\nsp\grampercc$, which is the region encountered by post--main-sequence massive stars \citep[see][and references therein]{Woosley2002The-evolution-a}. Stellar evolutionary calculations, which include the shock-induced explosive nucleosynthesis, find that a decrease in the $\carbon+\carbon$ rate leads  to enhancements in \aluminum[26] and \iron[60] abundances \citep{Gasques2007Implications-of}. Further calculations are needed to determine whether an enhanced $\carbon+\carbon$ rate would produce interesting changes in nucleosynthesis.

\subsection{Ignition in a Multicomponent Plasma}

In \S \ref{s.LMRdeviations} we showed that heavy spectator ions increase the plasma screening enhancement to the thermonuclear reaction rate via linear mixing rule deviations.  To our knowledge, this work is the first to show that spectator ions affect the plasma screening enhancement in the thermonuclear regime.  Prior work on the effects such deviations have on the plasma screening enhancement focused on binary ionic mixtures consisting only of reactants and products (e.g. \carbon\ and \magnesium) but no spectator ions \citep[e.g.][]{ogata.iyetomi.ea:equations,DSC96}, and \citet{DS03} concluded that linear mixing rule deviations always decrease the plasma screening enhancement in binary ionic mixtures (eq.~[\ref{e.H0BIM}]).  However, determining the effect of spectator ions requires analyzing a mixture of three or more ions, but previous applications of linear mixing rule deviations in ternary ionic mixtures \citep[e.~g.,][]{ogata.iyetomi.ea:equations} focused only on phase diagrams of crystallizing white dwarfs \citep[e.~g.,][]{IHMGB91,S96}, not on fusion reactions.

In our analysis, we tacitly assumed the plasma is uniformly mixed.  However, recent molecular dynamics simulations of multicomponent plasmas exhibit clustering of low-$Z$ ions \citep{WHSG08,Horowitz2008Thermal-conduct}, which may enhance the reaction rate.  This is worthy of further investigation.

For completeness, we note that linear mixing rule deviations are much larger for Coulomb solids \citep{DS03}.  This has two consequences for accreting neutron stars: (1) Screening enhancements for multicomponent plasmas in the pycnonuclear regime may be orders of magnitude greater than currently thought.  This could lower the pressures at which pycnonuclear reactions occur in the crust and thereby heat the ocean to a larger extent.  (2) A multicomponent plasma's freezing temperature is much lower than that of a one-component plasma.  This possibly explains the results of \citet{Horowitz2007Phase-Separatio}, whose molecular dynamics simulation of a multicomponent plasma froze at $\Gamma \approx 247$ rather than the typical $\Gamma \approx 175$ of a one-component plasma.

\subsection{Outlook for Future Measurements}\label{s.outlook}

Our conclusions are contingent on the existence of a strong resonance near $1.5 \nsp \MeV$ in the \carbon\ + \carbon\ system.  As noted in \S \ref{s.sfactor}, resonances exist throughout the experimentally studied energy range and are spaced at intervals of $\approx 0.3 \nsp \MeV$.  Therefore, a resonance almost certainly exists sufficiently near $1.5 \nsp \MeV$ (i.e., within the Gamow window).  We cannot predict with confidence, however, that the resonance strength $\og$ is sufficiently large.  Indeed, the measured resonances at higher energies typically increase the thermally averaged reaction rate $\sigmav$ by a factor $\lesssim 10$ over the nonresonant contribution, so the resonance needs to be unusually strong.  Resolving this issue requires experimental measurements of the \carbon\ + \carbon\ cross-section near the Gamow peak, which requires lab energies of $3\nsp\MeV$ with the ability to measure cross-sections at the $0.1\nsp\pico\barn$ level. Measurements at higher energies are
probably required to map out the resonance structure between the 
Gamow peak and the other available measurements of the fusion cross-section. Such measurements are possible in the near term with existing laboratories and are certainly within reach of planned underground facilities such as DUSEL \citep[though they will require the larger DUSEL accelerator option;][]{Goerres06}. 

\acknowledgments 
We thank Lars Bildsten, Philip Chang, Andrey Chugunov, Richard Cyburt,
Daniel Kasen, Hendrik Schatz, Michael Wiescher, Dima Yakovlev, Remco Zegers, and the anonymous referee for their advice and feedback.  The Joint Institute for Nuclear Astrophysics (\emph{JINA}) supported this work under NSF-PFC grant PHY~02-16783.  AWS and EFB are supported by NASA/ATFP grant NNX08AG76G, and RLC is supported by the National Science Foundation under Grant No.\ NSF PHY05-51164.

\begin{appendix}
\section{Derivation of $\Delta H(0)$}

In this Appendix, we derive $\Delta H(0)$, the correction to the plasma screening enhancement factor due to linear mixing rule deviations.  Consider a strongly-coupled Coulomb plasma consisting of $N \equiv \sum_{i} N_i$ ions, where $N_i$ is the number of ions with charge $Z_i$, $x_i \equiv N_i / N$ is the number fraction of species $i$, and species $1$ and $2$ are the reactant and product of the reaction, respectively, so that $Z_2 = 2 Z_1$.  \citet{dewitt.graboske.ea:screening} found that, neglecting quantum contributions, $H(0)$ equals the difference in the Coulomb free energy before and after the reaction; since a fusion reaction destroys two reactant ions and creates one product ion,
\begin{equation}\label{e.H0specific}
H(0) = F^{\mathrm{ex, initial}} - F^{\mathrm{ex, final}} =  \Fex(N_1, N_2, ..., N_n) - \Fex(N_1-2, N_2+1, ..., N_n).
\end{equation}
Writing equation (\ref{e.H0specific}) in a more general form, 
\begin{equation}\label{e.H0derivative}
H(0) = - \frac{\Fex(N_1-2\Delta N_2, N_2+\Delta N_2, ..., N_n) - \Fex(N_1, N_2, ..., N_n) }{\Delta N_2},
\end{equation}
where $\Delta N_2$ is the number of products created.  In the limit $\Delta N_2 \ll N_{1}, N_{2}$, equation (\ref{e.H0derivative}) simplifies to
\begin{equation}
H(0)= -\left (\pd{}{N_2} - 2\pd{}{N_1}\right ) \Fex.
\end{equation}
Using the general expression for $\Fex$ (eq.~[\ref{e.Fex2}]), we find
\begin{equation}\label{e.H0general}
\frac{H(0)}{\kB T}  = 2\fex(\Gamma_1) - \fex(\Gamma_2) +  \left \{\delfex - \dd \delfex \right \},
\end{equation}
where the term $2\fex(\Gamma_1) - \fex(\Gamma_2)$ is the well-known result for a plasma obeying the linear mixing rule \citep[e.g.,][]{J77}, the bracketed term is $\Delta H(0)/\kB T$, and we have defined the operator 
\begin{equation}\label{e.operator}
\dd \equiv  N \left (\pd{}{N_2} - 2\pd{}{N_1} \right ) = \pd{}{x_2} -2 \pd{}{x_1} + \displaystyle\sum_{i=1}^{n} x_i \pd{}{x_i}.
\end{equation}  

\citet{PCR09} derived an accurate, analytic fitting formula for $\delfex$ (see their eq.~[16]).  Using their formula for an unpolarized electron background, which is appropriate for a strongly-coupled Coulomb liquid, $\delfex = \delfex \left (\Gamma, \avgz, \avgzsq, \avgzfh \right )$, where $\Gamma = \sum_{i} x_i \Gamma_i$ and $\avgzk = \sum_{i} x_i Z_{i}^{k}$.  From equations (\ref{e.H0general}) and (\ref{e.operator}) of this work and equations (12), (14), and (16) of \citet{PCR09}, we find 
\begin{equation}
\frac{\Delta H(0)}{\kB T} = \delfex \left \{ 1 - \left [\pd{\ln \delfex}{\ln \Gamma} \dd \ln \Gamma + \pd{\ln \delfex}{\ln \avgz} \dd \ln \avgz + \pd{\ln \delfex}{\ln \avgzsq}  \dd \ln \avgzsq +\pd{\ln \delfex}{\ln \avgzfh} \dd \ln \avgzfh  \right ] \right \},
\end{equation}
where 
\begin{eqnarray}
\pd{\ln \delfex}{\ln \Gamma} &=& -\frac{abc \Gamma^{b}}{1+a \Gamma^{b}}, \quad \quad
\dd \ln \Gamma = 1 + (2^{5/3}-2)\frac{\Gamma_{1}}{\Gamma}, \quad  \quad \dd \ln \avgzk = 1 + (2^{k}-2) \frac{Z_1^{k}}{\avgzk}, \nonumber \\ \nonumber \\
\pd{\ln \delfex}{\ln \avgz} &=& -\frac{1}{2}\frac{\zetadh}{\zetadh-\zetalm} - \frac{ac \Gamma^{b}}{1+a \Gamma^{b}} \left [ \frac{1.1+34\delta^{3}}{2.2 \delta + 17 \delta^{4}} (1-\delta) + 0.4b \left(\ln \Gamma + \frac{1}{1-b} \right ) \right ] + \frac{d}{3} \ln(1+a\Gamma^{b}), \nonumber \\ \nonumber \\
\pd{\ln \delfex}{\ln \avgzsq} &=& \frac{3}{2}\frac{\zetadh}{\zetadh-\zetalm} + \frac{ac \Gamma^{b}}{1+a \Gamma^{b}} \left [ \frac{3.3+102\delta^{3}}{2.2 \delta + 17 \delta^{4}} (1-\delta) + 0.2b \left(\ln \Gamma + \frac{1}{1-b} \right ) \right ] - \frac{d}{6} \ln(1+a\Gamma^{b}), \nonumber \\ \nonumber \\
\pd{\ln \delfex}{\ln \avgzfh} &=& -\frac{\zetalm}{\zetadh-\zetalm} - \frac{ac \Gamma^{b}}{1+a \Gamma^{b}} \left [ \frac{2.2+68\delta^{3}}{2.2 \delta + 17 \delta^{4}} (1-\delta) \right ].
\end{eqnarray}

\end{appendix}

\bibliographystyle{apj}


\end{document}